\documentclass{article}
\usepackage{psfig}
\title{``Pressure Equilibration'' in Ultrarelativistic Heavy Ion Collisions$^B$}
\author{J. Brachmann, A. Dumitru, C. Spieles, \\
J. Maruhn, H. St\"ocker, W. Greiner,
 Univ.\ Frankfurt a.M.}
\date{}
\begin{document}
\maketitle

\noindent Almost 30 years ago hydrodynamics was used for the first time to 
describe heavy ion collisions at BEVALAC energies. 
The success of the macroscopic description of the collision dynamics 
indicated the creation of a hot and dense state of nuclear matter.
The great advantage of such an approach is that the equation of state (EoS) 
of nuclear matter 
enters directly into the hydrodynamic equations and it is
easy to implement phase transitions.
This success motivated the application of (one-fluid) hydrodynamics also 
for higher impact energies. The assumption of instantaneous
thermalization of projectile and target matter in this model leads to 
a maximum deposition of energy in the central reaction zone, so that already at
AGS energies the phase transition to a quark-gluon plasma (QGP) could be 
reached. This, for instance, leads to 
the breakdown of the directed flow in the reaction plane
since the pressure is not increasing anymore while the system is in 
the mixed phase \cite{Yaris}.
 
However, the assumption of instantaneous thermalization becomes unrealistic
when the rapidity gap of projectile and target is large.
Considering pp collisions at 24 AGeV \cite{refppdndy} 
the protons loose only one unit in rapidity and can be identified 
even after the collision.
It takes the nucleons several collisions to reach thermal equilibrium.
Moreover, the produced particles are also separated in phasespace.

In our hydrodynamical model,
we therefore consider three different fluids \cite{AD,3fluid} corresponding to
projectile and target nucleons and the produced particles, the so-called
{\em fireball}.
We assume local thermal equilibrium within each fluid but not among
different fluids.
Due to the above described forward-backward peaking of the pp-cross section
nucleons are not allowed to leave their individual fluids, 
so that the fireball remains net baryon free.
The coupling of the two nucleonic fluids is parametrized assuming free 
binary NN-collisions \cite{Satarov}. 
While penetrating each other, projectile and target fluid loose energy 
to the fireball.
The recoupling of the nucleonic fluids with the fireball is neglected here.
The fluids are propagated separately in three spacial dimensions so that
investigations with different impact parameters are possible.
When the local thermal velocities become comparable to the relative 
velocities of the (nucleonic) fluids they are merged into one.
For the nucleons we use the EoS of an ideal nonrelativistic gas with 
compressional energy.
Only the fireball may undergo a phase transition.

\begin{figure}[htb]
\centerline{\psfig{file=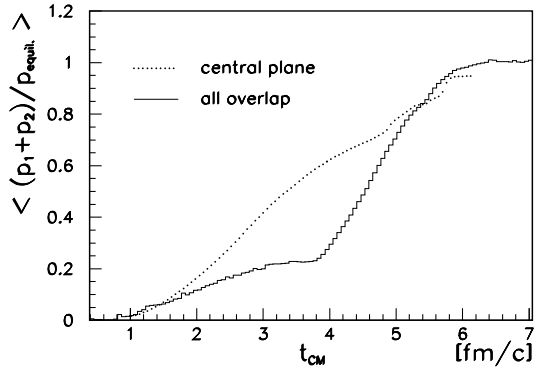,width=7.cm}}
\caption{Total pressure of the nucleonic fluids divided 
by equilibrium pressure (Au(11AGeV)+Au, b=0).}
\label{hydro_press}
\end{figure}
\begin{figure}[htb]
\centerline{\psfig{file=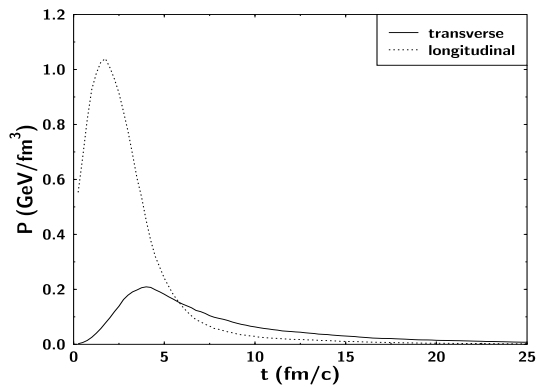,width=7.cm,height=5.cm}}
\caption{Longitudinal and transverse pressure in the central region of 
Au(11AGeV)+Au, b=0 in UrQMD.}
\label{urqmd_press}
\end{figure}
In Fig.\ \ref{hydro_press} we show the sum of the individual pressures 
of the nucleonic fluids divided by the corresponding equilibrium pressure,
i.e.\ the pressure of the fluids if they were equilibrated.
This ratio is averaged over the central plane perpendicular to the beam axis 
or over the overlap volume of the nuclei, respectively.
One observes that it takes $\approx 5~{\rm fm}/c=2R_{Au}/\gamma_{CM}$ 
to reach the equilibrium pressure.
A similar result is obtained within a microscopic cascade model (UrQMD,
\cite{Bass}).
Until $\approx 5$fm/c the longitudinal pressure is much larger 
than the transverse one (Fig.\ \ref{urqmd_press}). The pressures are averaged 
over a cylindrical volume of transverse radius $6$~fm and of $2$~fm length.
As a consequence of these nonequilibrium effects, even without a first order
phase transition, the directed nucleon flow
is considerably lower as compared to models where instant equilibration is
assumed (e.g.\ one-fluid hydrodynamics)\cite{3fluid}.  

\end{document}